\documentclass[12pt,preprint]{aastex}
\usepackage{lscape}

\usepackage{amsmath}

\newcommand{\lya}{Ly-$\alpha$ }

\newcommand{\lb}{Lyman break }
\newcommand{\zzz}{$z\sim 3$ }
\shorttitle{Candidate $z=3.03$ Brightest Proto-Cluster Galaxy} 
\shortauthors{Cooke et al.} 

\begin{document}

\title{A Candidate Brightest Proto-Cluster Galaxy at $z=3.03$}

\author{Jeff Cooke\altaffilmark{1} \altaffilmark{2}, Elizabeth
J. Barton\altaffilmark{1}, James S. Bullock\altaffilmark{1}, and Kyle
R. Stewart\altaffilmark{1}} \affil{The Center for Cosmology and the
Department of Physics \& Astronomy, The University of California,
Irvine, Irvine, CA, 92697-4575} \email{cooke@uci.edu, ebarton@uci.edu,
bullock@uci.edu, stewartk@uci.edu}

\author{and}
  
\author{Arthur M. Wolfe\altaffilmark{1}} \affil{Center for
Astrophysics and Space Sciences and the Department of Physics, The
University of California, San Diego, La Jolla, CA, 92093-0424}
\email{awolfe@ucsd.edu}

\altaffiltext{1}{Visiting Astronomer, W. M. Keck Telescope. The Keck
  Observatory is a joint facility of the University of California, the
  California Institute of Technology, and NASA and was made possible
  by the generous financial support of the W. M. Keck Foundation.}
  \altaffiltext{2}{Gary McCue Postdoctoral Fellow}


\begin{abstract}
 
We report the discovery of a very bright (m$_R=22.2$) Lyman break
galaxy at $z=3.03$ that appears to be a massive system in a late stage
of merging.  Deep imaging reveals multiple peaks in the brightness
profile with angular separations of $\sim$$0.''8$ ($\sim$25 $h^{-1}$
kpc comoving).  In addition, high signal-to-noise ratio rest-frame UV
spectroscopy shows evidence for $\sim5$ components based on stellar
photospheric and ISM absorption lines with a velocity dispersion of
$\sigma \sim 460$ km s$^{-1}$ for the three strongest components.
Both the dynamics and high luminosity, as well as our analysis of a
$\Lambda$CDM numerical simulation, suggest a very massive system with
halo mass $M \sim10^{13} M_\odot$.  The simulation finds that all
halos at $z=3$ of this mass contain sub-halos in agreement
with the properties of these observed components and that such systems
typically evolve into $M \sim 10^{14} M_{\odot}$ halos in
groups and clusters by $z=0$.  This discovery provides a rare
opportunity to study the properties and individual components of \zzz
systems that are likely to be the progenitors to brightest cluster
galaxies.

\end{abstract}



\section{Introduction}

Massive galaxies and galaxy clusters at high redshift provide strong
constraints on cosmological models and galaxy formation scenarios.
The \lb galaxies (LBGs) are a widespread population of high redshift
galaxies selected by their rest-frame FUV colors \citep{s96,s98}.
LBGs at $z\sim3$ tend to reside in massive halos $\langle M_{DM}
\rangle \sim 10^{11.5} M_\odot$ \citep{a05} and are therefore believed
to evolve into present-day massive elliptical galaxies.  Although
$\sim3000$ color-selected $z>2$ LBGs have spectroscopic confirmations
\citep[e.g.][]{s03,s04,c05,c08}, only two systems, selected by their
radio emission, have been spectroscopically identified as potential
brightest cluster galaxies \citep{zirm05,miley06}.

We report the discovery of a bright, m$_R=22.2$, rest-frame FUV
color-selected LBG (hereafter LBG-2377) at $z=3.03$.  Ground-based
imaging and spectroscopy finds no evidence for AGN activity or
gravitational lensing making this the most luminous $z\gtrsim3$ LBG
discovered to date.  The deep images and high signal-to-noise ratio
(SNR) spectroscopy show evidence for multiple massive components
indicative of a system in the late stage of merging.  We show that the
component velocity dispersion and integrated FUV luminosity suggest
that LBG-2377 resides in a very massive halo ($\sim10^{13} M_\odot$)
and is likely to evolve into a present-day massive brightest cluster
galaxy (\S~\ref{sim}).  Therefore, LBG-2377 provides a rare
opportunity to study the dynamics and properties of a massive LBG halo
in the process of formation.

All $z\gtrsim3$ LBGs with $R<23$ discovered to date are magnified by
gravitational lensing.  Because all of these show \lya in absorption,
they are representative of the spectral properties observed in
$\sim$25-50\% of the LBG population \citep{aes03}.  LBG-2377 displays
strong \lya in emission in two of its components and, because of its
brightness, is the only system of its kind.  The luminosity of
LBG-2377 enables high SNR study of its individual components and
thereby a means to constrain the observed range of LBG spectral
properties.  We briefly describe the imaging and spectroscopic 
observations of LBG-2377 in \S\ref{obs}, highlight the initial 
analysis of the data in \S\ref{analysis}, and present a discussion 
in \S\ref{disc}.  All magnitudes are in the AB \citep{f96} magnitude 
system unless otherwise noted and we assume an $\Omega_M=0.3$, 
$\Omega_\Lambda=0.7$ cosmology and use comoving units throughout.


\section{Observations}\label{obs}

\subsection{Imaging}\label{imaging}

The deep images of the QSO PC1643+4631A field were obtained on 2001
April 18 using the Low-Resolution Imaging Spectrometer \citep{o95}
mounted on the Keck I telescope as part of a multi-field survey for
LBGs associated with damped \lya systems at \zzz \citep{c05}.
Specifically, the field containing LBG-2377 was imaged in the
$u'$BVR$_S$\footnote{R$_s$ is the H. Spinrad high-throughout night-sky
R filter publicly available at the W. M. Keck Observatory.  The
transformation from Spinrad night-sky R$_S$ into Johnson VR is (R$_S$
- R) = -0.004 - 0.072(V - R) + 0.073(V - R)$^2$ \citep{stern99}.  All
R$_S$ magnitudes presented here have been transformed into the Johnson
R bandpass.}I filters with seeing of $0.7-0.9$ arcsec and a depth of
$u'=28.8$ and BVRI $\ge28.1$ ($1\sigma$).  We discovered LBG-2377 as
an R $=22.2$ $z\sim3$ LBG candidate at R.A. 16:44:48.3,
Dec.+46:27:08.2 (J2000.0) toward one edge of the imaged field.  The
isophotal magnitudes as determined using $SExtractor$ \citep{ba96}
source extraction code are: $u'=26.5 \pm0.46$, B $=24.3 \pm0.10$, V
$=22.6 \pm0.03$, R $=22.2 \pm0.06$, and I $=22.3 \pm0.07$.

\begin{figure}
\begin{center}
\scalebox{0.6}[0.7]{\rotatebox{90}{\includegraphics{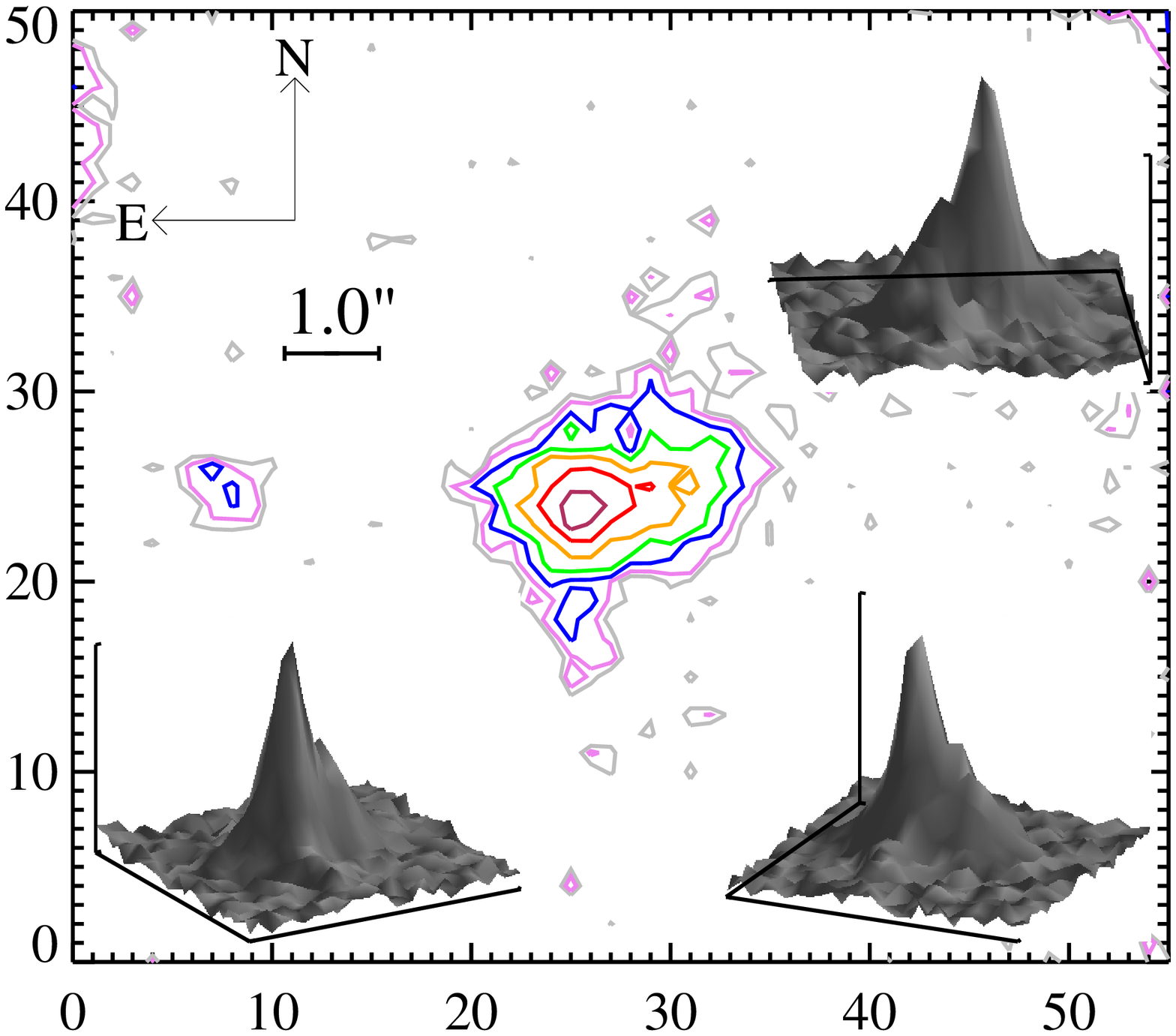}}}
\caption
{Contour plots of LBG-2377.  The 2-D plot indicates a primary peak
  located slightly to the E of the object centroid.  Secondary and
  tertiary peaks are detected W and SE of the centroid, respectively.
  Three 3-D contour plots are inset with rotations (clockwise from
  upper-right), $185^\circ$, $335^\circ$, and $30^\circ$
  (S$=0^\circ$).}
\label{fig:contour}
\end{center}
\end{figure}

LBGs at \zzz have a half-light radius of $\lesssim0.''3$
\citep[e.g.,][]{g01}.  As illustrated in the contour plot shown in
Figure~\ref{fig:contour}, LBG-2377 spans $\sim2''\times3''$ and has a
morphology that can be interpreted as a single extended system
exhibiting multiple vigorous star forming regions or multiple separate
near-point sources.  The ratio of the peaks of the three strongest
identified components is approximately 10:5:1 with separations of
$\sim$$0.''6-1''.0$ ($\sim20-30h^{-1}$ kpc, comoving).  Inspection of
the deep images shows that there are no significant lensing-source
candidates within $\sim10''$.  A potential candidate $20''$ from
LBG-2377 (R.A. 16:44:46.5, Dec +46:27:04.1) does have an effective
redshift for lensing ($z_{phot}=0.41\pm0.2$), but would need a halo
mass of $M\gtrsim7\times10^{14} M_\odot$ to provide a significant
($\gtrsim2-3\times$) magnification boost \citep{minor07}.  Moreover,
such a geometry would result in a large distortion of the LBG-2377
image.  We see no evidence for this in the deep ground-based images.

\subsection{Spectroscopy}\label{spectroscopy}

\begin{figure}
\begin{center}
\scalebox{0.75}[0.65]{\rotatebox{90}{\includegraphics{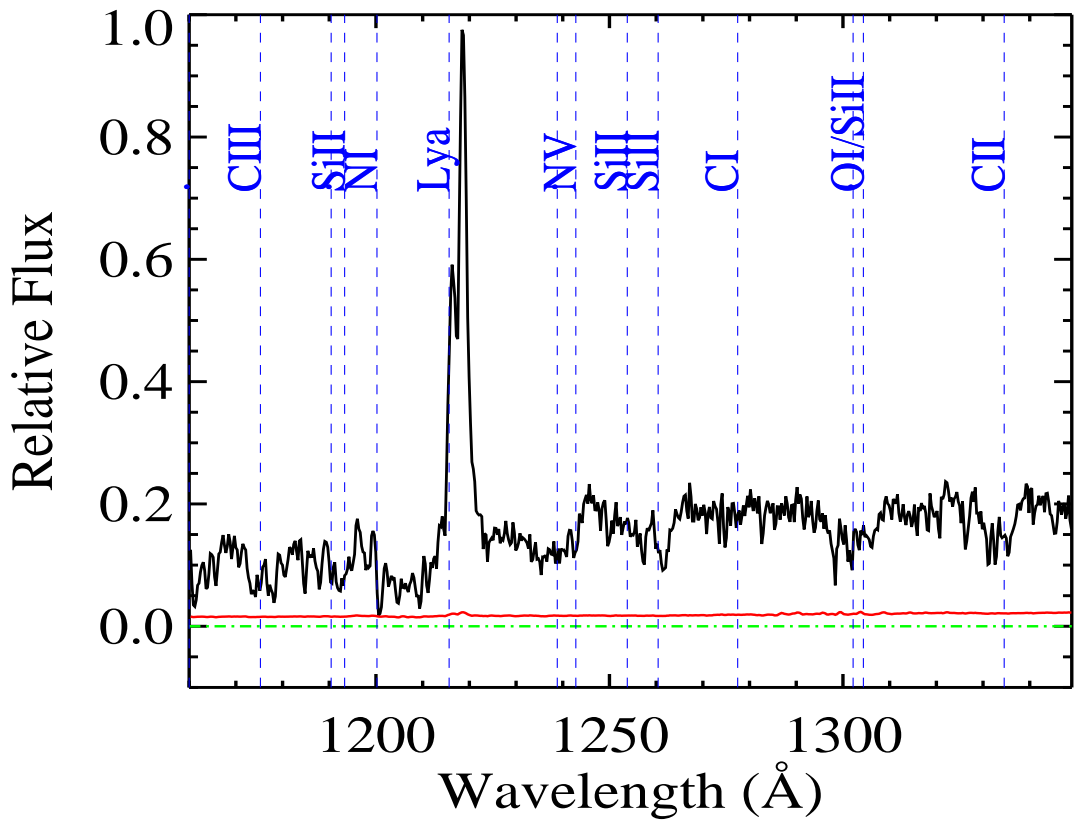}}}
\caption
{Spectrum of LBG-2377.  This section is shown to illustrate both the
 double-peak \lya emission feature and broad, complex interstellar
 absorption features [vertical {\it (blue)} dashed lines].  In all
 sepctra in this paper, the green line indicates zero flux and the red
 line is the error array.}
\label{fig:600}
\end{center}
\end{figure}

LBG-2377 was confirmed to be a $z=3.03$ galaxy from low-resolution
discovery multi-object spectroscopy on 2004 February 18 \citep{c05}.
The data were taken using the 300 line mm$^{-1}$ grism on LRIS with a
spectral resolution of $\sim10$\AA~ FWHM.  We acquired follow-up
higher resolution, higher SNR longslit spectroscopic observations with
LRIS on 21 May 2007 using the 600 line mm$^{-1}$ grism blazed at
4000\AA~with the blue arm and the 1200 line mm$^{-1}$ blazed at 7500
\AA~with the red arm.  We used five 1800s exposures in our final
combined spectrum that resulted in a SNR of $10-15$ and a resolution
of $\sim2-3$\AA (Figure~\ref{fig:600}).  Details of the observations
will appear in a forthcoming paper.

The spectrum of LBG-2377 is dominated by O and B star continua and
displays strong \lya emission and little reddening.  These features
are consistent with the general properties of the $z\sim3$ LBG
population that display \lya in emission \citep{aes03,c05}. Although
the \lya feature appears as a single emission line in the
low-resolution discovery spectrum, the higher-resolution high SNR
spectroscopy reveals two strong peaks with two possible weak peaks
more evident in the individual spectra.  The rest-frame FUV shows a
complex series of interstellar atomic lines exhibiting gas absorption
over velocities $\gtrsim2000$ km s$^{-1}$ and evidence for $\sim5$
components.  Over 20 FUV ISM absorption lines were used to identify
and study individual components.


\section{Analysis}\label{analysis}

\subsection{Components}\label{velocities}

We identified $\ge$5 photospheric lines (to varying confidence levels)
of each component to estimate their systemic redshifts.  These lines
include\footnote{For consistency with future analyses and because our
observations are of similar quality, we use the stellar photospheric
values listed in Table 1 of \citet{pet00}.}: Si\textsc{iii}
$\lambda1294.54$, 1296.73, 1417.24, C\textsc{iii} $\lambda1175.26,
1296.33$, 1427.85 C\textsc{ii} $\lambda$1323.93, N\textsc{iii}
$\lambda1324.32$, O\textsc{iv} $\lambda1343.35$, and S\textsc{v}
$\lambda1501.76$.  The strongest components have redshifts $z=3.0385,
3.0289$, and 3.0244.  We find a 1-D velocity dispersion of $\sigma =
460$ km s$^{-1}$ for these three components and $\sigma = 705$ km
s$^{-1}$ for the five components.  The details of the identified 
components are listed in Table~\ref{comp}.  

\begin{figure}
\begin{center}
\scalebox{0.35}[0.40]{\rotatebox{90}{\includegraphics{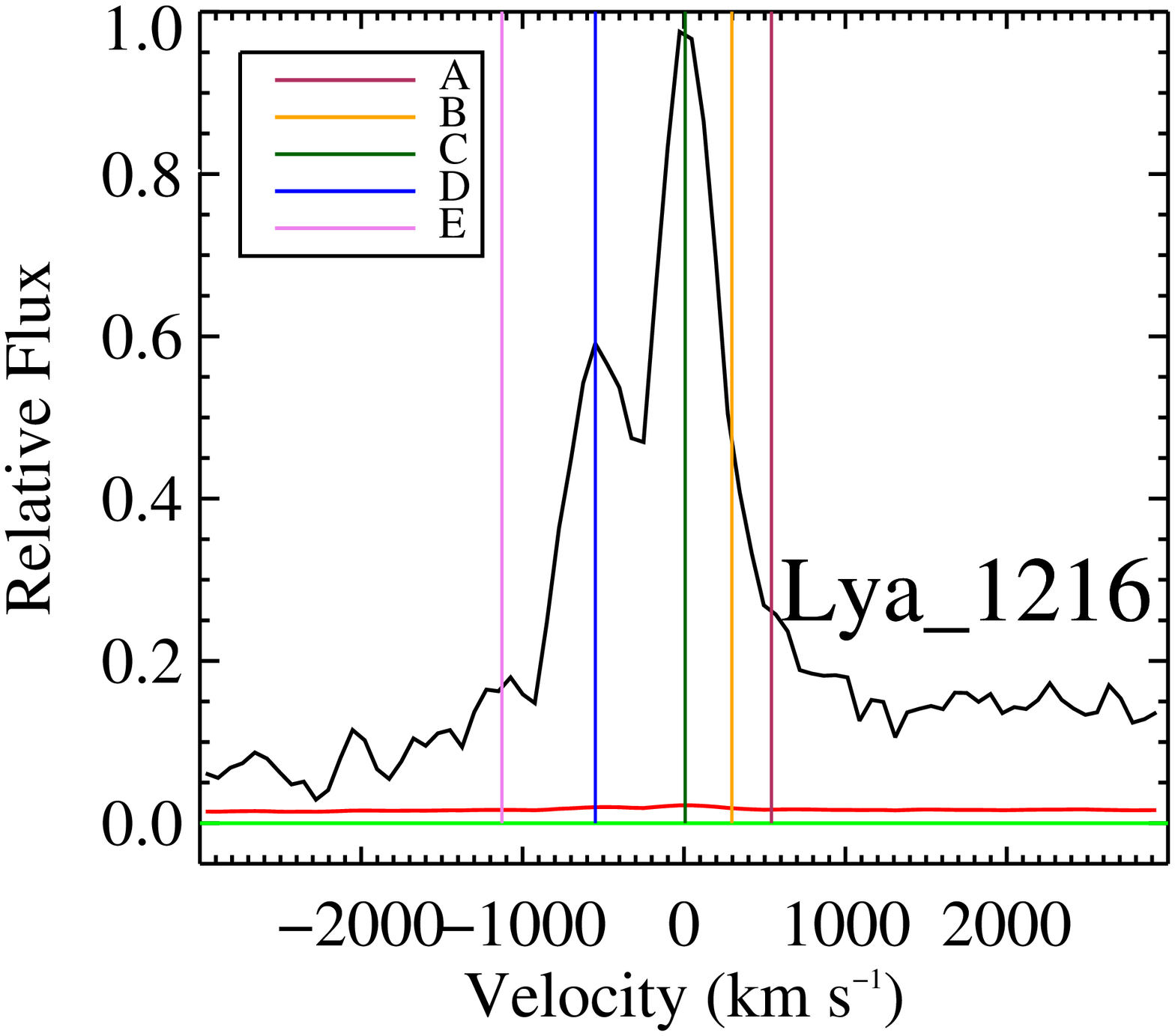}}}
\scalebox{0.35}[0.40]{\rotatebox{90}{\includegraphics{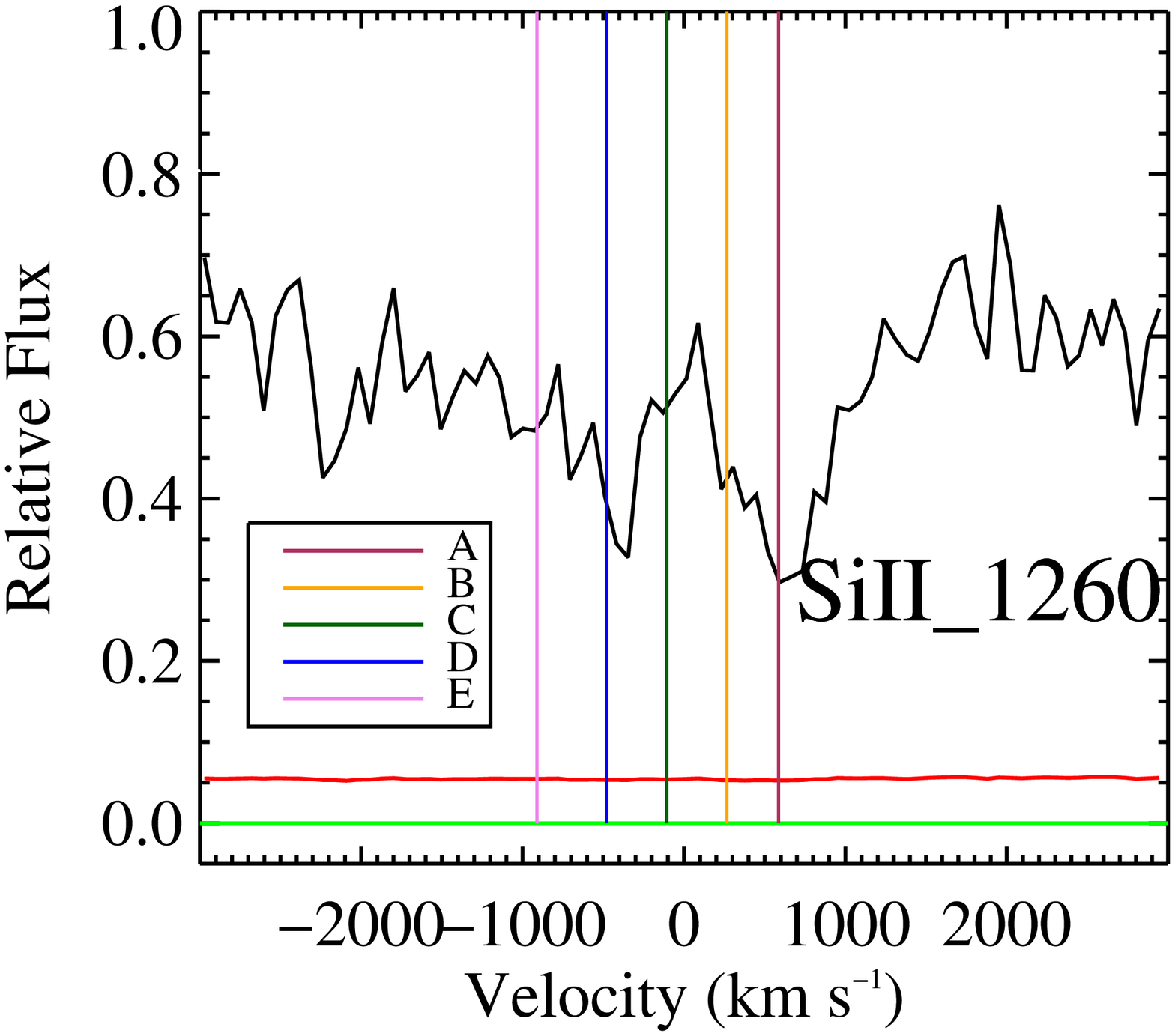}}}
\scalebox{0.35}[0.40]{\rotatebox{90}{\includegraphics{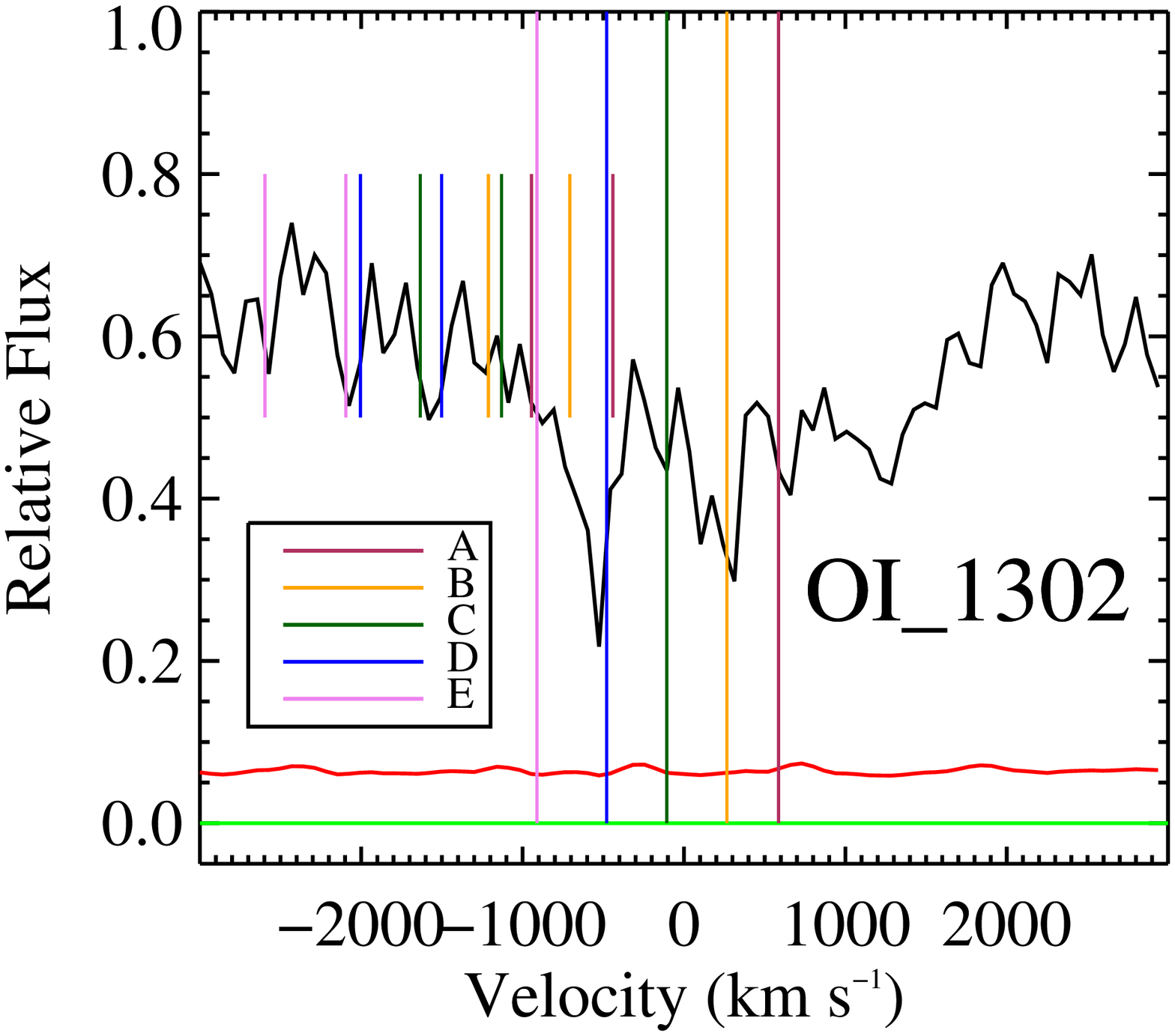}}}
\scalebox{0.35}[0.40]{\rotatebox{90}{\includegraphics{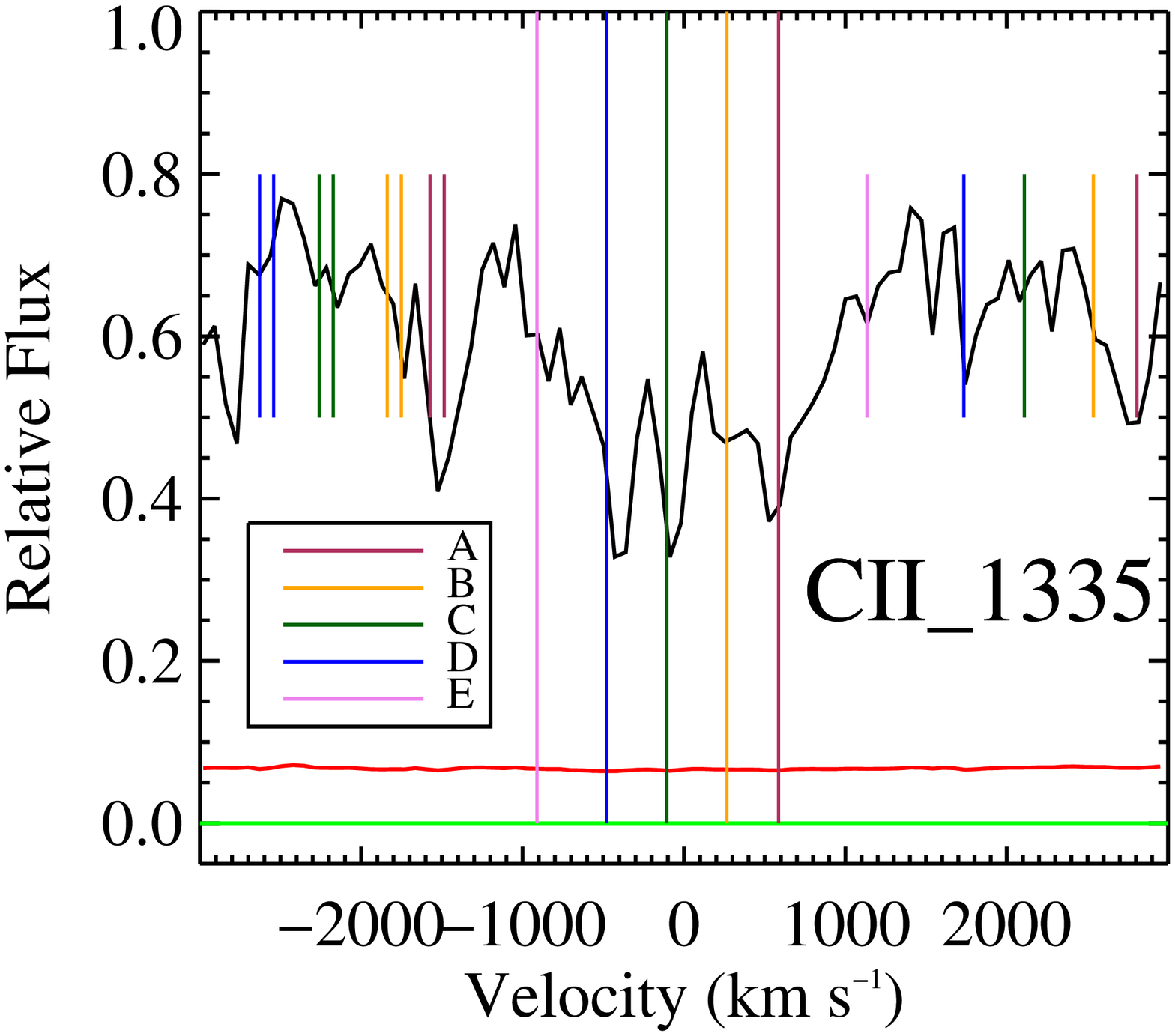}}}
\caption
{Spectrum of LBG-2377 showing five tentative components (A - E in
  Table~\ref{comp}) identified using 21 ISM features.  Velocities.
  are with respect to the average of components A, C, \& D.  {\it
  Upper left:} The two strong \lya emission lines (components C \& D)
  and three potential weak \lya lines [A, B, \& E].  {\it Clockwise
  from the upper right:} Fits for the ISM transitions Si\textsc{ii}, 
  C\textsc{ii}, and O\textsc{i}.  Also shown are the stellar 
  photospheric features Si\textsc{iii} 1294.5, 1296.7, C\textsc{iii} 
  1296.3, C\textsc{ii} 1323.9, N\textsc{iii} 1324.3, and O\textsc{iv} 
  1343.4 (not labeled for clarity) indicated by short vertical lines.  
  The signal-to-noise ratio is $\sim15$ in these regions, therefore the 
  more prominant features are real.  Each absorption feature is in 
  agreement with their respective systemic redshift to within the 
  spectral resolution ($\sim150$ km sec$^{-1}$).}
\label{fig:spectra}
\end{center}
\end{figure}

We note that the \lya emission and ISM absorption lines of the LBG
population show observed velocity offsets of $\sim+650$ and $\sim-200$
km sec$^{-1}$ from their systemic velocities, respectively
\citep{pet02,a03,c05}.  The \lya and ISM offsets in the identified 
components are consistent with that of the LBG population.  The 
accepted interpretation is that these offsets are caused by the 
presence of an expanding galactic-scale shell of gas and dust driven 
by supernovae and stellar winds.  \lya photons are largely absorbed in 
the approaching shell but are resonantly scattered off the receding 
portion and survive.  In this scenario, the two prominent \lya peaks 
observed in the spectrum of LBG-2377 would trace two distinct 
expanding galactic-scale shells with a systemic velocity difference 
of $\sim550$ km sec$^{-1}$.  Our fit to the components in 
Table~\ref{comp} is shown in Figure~\ref{fig:spectra}.  Although the 
spectra of the components are superposed, the ISM absorption-line 
strengths and behavior of each component appear internally consistent 
with being separate systems.

\subsection{Theoretical expectations}\label{sim}

The high luminosity and complex structure of LBG-2377 suggest that
this system may be a brightest cluster galaxy in formation.  Here we
investigate this possibility in the context of $\Lambda$CDM.  We
estimate the dark matter halo mass of LBG-2377 using both its measured
kinematics and luminosity.  Encouragingly, both estimates suggest a
group-scale mass $\sim 10^{13} M_\odot$.  If we assume an NFW halo,
our measured $\sigma \sim 460$ km s$^{-1}$ 1-D velocity dispersion
corresponds to a halo maximum circular velocity of $V_{\rm max} \simeq
1.5 \sigma$, or $V_{\rm max} \sim 700$ km s$^{-1}$ \citep{klypin99}.
At $z \sim 3$ this implies $M_{\rm vir} \sim 10^{13} \, M_\odot$
\citep[e.g.][]{bullock01}.  Alternatively, if we assume a simple
monotonic relationship between galaxy luminosity and dark matter halo
$V_{\rm max}$ \citep[e.g.][]{berrier06} and use the \zzz LBG
luminosity function \citep{s99,saw06}, the number density of objects
similar to LBG-2377 (m$_R=22.2$) is $\sim 10^{-7}$ Mpc$^{-3}
h^{3}$.  This number density corresponds to halos with $V_{\rm max}
\sim700$ km s$^{-1}$ at $z=3$ \citep{berrier06}.  The agreement
between these two estimates suggests that LBG-2377 does indeed
sit within a massive $V_{\rm max} \sim 700$ km s$^{-1}$ halo.

Halos that are this massive at $z=3$ almost certainly contain
substructure.  We illustrate this explicitly by examining the
high-resolution 80 $h^{-3}$ Mpc$^{-3}$ $\Lambda$CDM N-body simulation
described in \cite{stewart07}.  There are three halos in this box with
$V_{\rm max} > 600$ km s$^{-1}$ at $z=3$ (specifically, $V_{\rm max} =
620, 645,$ and $740$ km sec$^{-1}$).  These objects have between five
and eight $10^{11} h^{-1} M_\odot < M < 10^{12.5} h^{-1} M_\odot$
sub-halos within their (physical) virial radii $R_{\rm vir} \simeq 90
h^{-1}$ kpc $(M/10^{13} h^{-1} M_\odot)^{1/3}$.  Note that this
substructure mass is typical of that expected for LBG hosts at \zzz
\citep[$\langle M_{LBG}\rangle \sim$10$^{11.5} M_\odot$][]{a05}.  If
we look at simulated halos that are significantly smaller than our
estimate, we find that only two out of 16 objects with $V_{\rm max} >
450$ km sec$^{-1}$ in the simulation have no sub-halos with $M >
10^{11} h^{-1} M_\odot$.  We conclude that it would be very unlikely
for an object like LBG-2377 to exist {\em without} multiple
components.

By following the merger trees of our three $V_{max} \gtrsim 600$ km
sec$^{-1}$ halos, we find that they evolve from $M=10^{13.0-13.5}
h^{-1} M_\odot$ halos at $z=3$ to cluster-mass systems with
$M=10^{13.9-14.3} h^{-1} M_\odot$ at $z=0$.  These are among the $12$
most massive halos in the box at $z=0$ and contain $\sim 10-50$
sub-halos larger than $10^{11} h^{-1} M_\odot$.  This analysis
suggests that LBG-2377 will evolve into the dominant galaxy within a
large group or cluster by the present day.

\subsection{\lya emission}

The double peak nature of the \lya feature in LBG-2377 was only
detected with higher resolution and SNR.  However, multiple \lya peaks
are not unique to LBG-2377.  In the LBG spectroscopic survey of
\citet{c05}, we find $\sim3$\% of the low-resolution spectra to
display double \lya emission peaks and resolved double peaks in their
deep image profiles.  We can only assume that this fraction increases
with higher resolution or SNR.  One example of this effect may be
found in the sample of \citet{aes06}.  Of the nine deep high SNR LBG
spectra with \lya in emission, three have double peaks.  Two of the
remaining six have redshifts within $\Delta z = 0.01$ and a separation
of $\sim2''$ and are consistent with a double-peak system at a larger
angular separation.  Interestingly, upon inspection of the average
internal velocity differences between the \lya emission and ISM
absorption features ($\Delta z = z_{em} - z_{abs}$), the single peak
systems are reported to have $\langle\Delta z\rangle \sim$550 km
sec$^{-1}$, whereas the double peak systems have $\langle\Delta
z\rangle \sim1200$ km sec$^{-1}$
\footnote{This value omits the system C32 that shows a higher redshift
for absorption than emission for one of the two components.  Including
this system results in a $\Delta z \sim900$ km sec$^{-1}$.  If the
absorption features of the lower redshift systems are weak or narrow
in this interpretation, adopting a typical $\Delta z$ for the that
system would result in $\Delta z \sim1100$ km sec$^{-1}$ for the
assumed two components.}. One plausible interpretation of this
discrepancy is the result of the velocity separations, and consequent
spectral blending, of two individual interacting systems.  Confirming
separate components in LBGs showing only \lya in absorption is more
difficult.  However, there are indications of separate components in
the intermediate-resolution, high SNR spectrum of MS1512-cB58
\citep{pet02}.  The prediction implied is that with higher SNR and/or
resolution, a significant fraction, possibly as high as $\sim30$\%, of
LBGs with \lya in emission have multiple massive components and may
indeed be separate interacting or merging systems. This result would
be consistent with the fraction of expected major mergers at $z\sim3$
from numerical simulations \citep[e.g.][] {wechsler01,ryan07}), and
with galaxy formation theories that suggest mergers are an important
contributor to the star formation activity in LBGs
\citep{kolatt99,somerville01}.


\section{Discussion}\label{disc}

What is the nature of LBG-2377?  Possible interpretations are that
LBG-2377 represents either (1) a single extended LBG with $M
\sim10^{13}M_\odot$ exhibiting multiple regions of vigorous star
formation, (2) a single, extended LBG with mass more typical of LBGs
($M\sim10^{12}M_\odot$) that appears over-luminous due to
exceptionally high bursts of star formation, (3) multiple unbound $M
\sim10^{12}M_\odot$ LBGs in near proximity, or (4) multiple LBGs in a
$M \sim10^{13}M_\odot$ halo observed in a late stage of merging.  We
discuss each scenario below and show that the evidence gathered to
date favors interpretation 4.

If the component stellar photospheric lines are misidentified and the
complex nature of the ISM absorption and \lya emission lines are
ascribed to violent mixing of the ISM and discrete star forming
regions separated by $\ge 25h^{-1}$ kpc, interpretation 1 may hold.
However, a single $M\sim10^{13}M_\odot$ halo with no massive sub-halos
is in conflict with simulation results.  In this interpretation,
LBG-2377 is indeed a massive and extended system that appears to be
experiencing a monolithic burst of star formation.  Under
interpretation 2, LBG-2377 would be in less conflict with simulations
yet the above assumptions for stellar and ISM lines must hold.  In
this case, LBG-2377 is a single system undergoing an unusually strong
star formation episode at a level that has not been witnessed in any
other LBG.  In addition, this exceptional star formation must be
occurring in multiple regions separated by $\ge 25h^{-1}$ kpc.  If the
velocity offsets are interpreted as redshifts, as in interpretation 3,
the systems would have separations of $\sim5h^{-1}$ Mpc comoving.  An
event of three or more unbound systems with $M\sim10^{12}M_\odot$,
$\le 1.''0$ angular separations, and an LBG density of
$\sim10^{-3}-10^{-4}h^{3}$ Mpc$^-3$ has a vanishingly low probability,
even if the systems were given reasonable peculiar velocities.  The
most likely scenario is interpretation 4 because it is consistent with
the extended nature and multiple peaks seen in the images, the
superposed spectra of multiple identified components, the observed
velocity dispersion, and the expectations from the simulation.  In
addition, this interpretation provides a natural driver for the
observed star formation and bright integrated luminosity.
Furthermore, the double \lya emission peak of LBG-2377 is also seen in
other LBGs which exhibit two distinct components in their deep images
and \lya emission-line velocity differences consistent with close
galaxy pairs.

LBG-2377 is a fortuitous discovery that enables detailed study of the
individual components of a candidate massive galaxy in the late stage
of merging.  Study of the relationships between the spectral and
spatial properties of this single system will constrain the formation
processes of the LBG population and lend insight into the fraction of
LBGs detected as a result of merger induced star formation.  Future
high-resolution ground-based IFU data on existing instruments, such as
OSIRIS on Keck, have the capability to resolve the velocity structure
of LBG-2377 and hence its true nature.


\acknowledgments

J.C. is supported in part by a Gary McCue postdoctoral fellowship
through the Center for Cosmology at the University of California,
Irvine.  E.J.B. and J.S.B. are supported by the Center for Cosmology at
the University of California, Irvine and J.S.B. and K.S. are supported
by NSF grant AST-0507816.  The authors wish to recognize and
acknowledge the very significant cultural role and reverence that the
summit of Mauna Kea has always had within the indigenous Hawaiian
community.  We are most fortunate to have the opportunity to conduct
observations from this mountain.


\begin{thebibliography}{}

\bibitem[Adelberger et al.(2003)]{a03} Adelberger, K. L., Steidel,
  C. C., Shapley, A. E., \& Pettini, M. 2003, ApJ, 584, 45
\bibitem[Adelberger et al.(2005)]{a05} Adelberger, K. L., Steidel,
  C. C., Pettini, M., Shapley, A. E., Reddy, N. A., \& Erb,
  D. K. 2005, ApJ, 619, 697
\bibitem[Berrier et al.(2006)]{berrier06} Berrier, J., Bullock, J. S.,
  Zentner, A. R., Guenther, H., Barton, E. J., Kravtsov, A. V., \&
  Wechsler, R. H. 2006, ApJ, 652, 56
\bibitem[Bertin \& Arnouts(1996)]{ba96} Bertin, E. \& Arnouts,
  S. 1996, A\&AS, 117, 393
\bibitem[Bullock et al.(2001)]{bullock01} Bullock, J.~S., Kolatt,
  T.~S., Sigad, Y., Somerville, R.~S., Kravtsov, A.~V., Klypin, A.~A.,
  Primack, J.~R., \& Dekel, A.\ 2001, \mnras, 321, 559
\bibitem[Cooke et al.(2005)]{c05} Cooke, J., Wolfe, A. M., Gawiser,
  E., \& Prochaska, J. X. 2005, ApJ, 621, 596
\bibitem[Cooke et al.(2008)]{c08} Cooke, J., Barton, E., Barbour,
  E. Gawiser, E., \& Wolfe, A. 2008 in preparation
\bibitem[Fukugita et al.(1996)]{f96} Fukugita, M., Ichikawa, T., Gunn,
  J. E., Doi, M., Shimasaku, K., \& Schneider, D. P. 1996, AJ, 111,
  1748
\bibitem[Gardner et al.(2001)]{g01} Gardner, J. P., Brown, T. M., \&
 Ferguson, H. C. 2000, ApJ, 542L, 79
\bibitem[Klypin et al.(1999)]{klypin99} Klypin, A.; Gottlöber, S.,
  Kravtsov, A. V., \& Khokhlov, A. M.  1999, ApJ, 516, 530
\bibitem[Kolatt et al.(1999)]{kolatt99} Kolatt, T. S., Bullock, J. S.,
  Somerville, R. S., Sigad, Y., Jonsson, P., Kravtsov, A. V., Klypin,
  A. A., Primack, J. R., Faber, S. M., \& Dekel, A. 1999, ApJ, 523,
  109
\bibitem[Law et al.(2007)]{law07} Law, D. R., Steidel, C. C., Erb,
  D. K., Pettini, M., Reddy, N. A., Shapley, A. E., Adelberger, K. L.,
  \& Simenc, D. J. 2007, ApJ, 656, 1
\bibitem[Miley et al.(2006)]{miley06} Miley, George K., Overzier,
  Roderik A., Zirm, Andrew W., Ford, Holland C., Kurk, Jaron,
  Pentericci, Laura, Blakeslee, John P., Franx, Marijn, Illingworth,
  Garth D., Postman, Marc, Rosati, Piero, R\"{o}ttgering, Huub J. A.,
  Venemans, Bram P., \& Helder, Eveline 2006, ApJ, 650, 29
\bibitem[Minor(2007)]{minor07} Minor, Q. 2007, private communication.
\bibitem[LRIS; Oke et al.(1995)]{o95} Oke, J. B., Cohen, J. G., Carr,
  M., Cromer, J., Dingizian, A., Harris, F. H., Labrecque, S.,
  Lucinio, R., Schaal, W., Epps, H., \& Miller, J. 1995, PASP, 107,
  375
\bibitem[Pettini et al.(2000)]{pet00} Pettini, M., Steidel, C. C.,
  Adelberger, K. L., Dickinson, M., \& Giavalisco, M. 2000, ApJ, 528,
  96
\bibitem[Pettini et al.(2002)]{pet02} Pettini, M., Rix, S. A.,
 Steidel, C. C., Adelberger, K. L., Hunt, M. P., \& Shapley,
 A. E. 2002, ApJ, 569, 742
\bibitem[Ryan et al.(2007)]{ryan07} Ryan, R. E., Cohen, S. H.,
  Windhorst, R. A., \& Silk, J. 2007, arXiv:0712.0416
\bibitem[Sawicki \& Thompson(2006)]{saw06} Sawicki, M. \& Thompson,
  D. 2006, ApJ, 642, 653
\bibitem[Shapley et al.(2003)]{aes03} Shapley, A. E., Steidel, C. C.,
  Adelberger, K. L., \& Pettini, M. 2003, ApJ, 588, 65
\bibitem[Shapley et al.(2006)]{aes06} Shapley, A. E., Steidel, C. C.,
  Pettini, M., Adelberger, K. L., \& Erb, D. K. 2006, ApJ, 651, 688
\bibitem[Somerville, Primack \& Faber(2001)]{somerville01} Somerville,
  R. S., Primack, J. R., \& Faber, S. M. 2001 MNRAS, 320, 504
\bibitem[Steidel et al.(1996)]{s96} Steidel, C. C., Giavalisco, M.,
  Pettini, M., Dickinson, M., \& Adelberger, K. L. 1996, ApJ, 462, 17
\bibitem[Steidel et al.(1998)]{s98} Steidel, C. C., Adelberger, K. L.,
  Giavalisco, M., Dickinson, M., Pettini, M., \& Kellogg, M. 1998,
  \apj, 492, 428
\bibitem[Steidel et al.(1999)]{s99} Steidel, C. C, Adelberger, K. L.,
  Giavalisco, M., Dickinson, M., \& Pettini, M. 1999, ApJ, 519, 1
\bibitem[Steidel et al.(2003)]{s03} Steidel, C. C., Adelberger, K. L.,
  Shapley, A. E., Pettini, M., Dickinson, M., \& Giavalisco, M. 2003,
  ApJ, 592, 728
\bibitem[Steidel et al.(2004)]{s04} Steidel, C. C., Shapley, A. E.,
  Pettini, M., Adelberger, K. L., Erb, D. K., Reddy, N. A., \& Hunt,
  M. P. 2004, ApJ, 604, 534
\bibitem[Stern et al.(1999)]{stern99} Stern, D., Dey, A., Spinrad, H.,
  Maxfield, L., Dickinson, M., Schlegel, D., \& González,
  R. A. 1999, AJ, 117, 1122
\bibitem[Stewart et al.(2007)]{stewart07} Stewart, K. R., Bullock,
  J. S., Wechsler, R. H., Maller, A. H., \& Zentner,
  A. R.2007, arXiv:0711.5027
\bibitem[Wechsler et al.(2001)]{wechsler01} Wechsler, R. H.,
  Somerville, R. S., Bullock, J. S., Kolatt, T. S., Primack, J. R.,
  Blumenthal, G. R., \& Dekel, A. 2001, ApJ, 554, 85
\bibitem[Zirm et al.(2005)]{zirm05} Zirm, A. W., Overzier, R. A.,
  Miley, G. K., Blakeslee, J. P., Clampin, M., De Breuck, C., Demarco,
  R., Ford, H. C., Hartig, G. F., Homeier, N., Illingworth, G. D.,
  Martel, A. R., R\"{o}ttgering, H. J. A., Venemans, B., Ardila,
  D. R., Bartko, F., Ben\'{i}tez, N., Bouwens, R. J., Bradley, L. D.,
  Broadhurst, T. J., Brown, R. A., Burrows, C. J., Cheng, E. S.,
  Cross, N. J. G., Feldman, P. D., Franx, M., Golimowski, D. A., Goto,
  T., Gronwall, C., Holden, B., Infante, L., Kimble, R. A., Krist,
  J. E., Lesser, M. P., Mei, S., Menanteau, F., Meurer, G. R., Motta,
  V., Postman, M., Rosati, P., Sirianni, M., Sparks, W. B., Tran,
  H. D., Tsvetanov, Z. I., White, R. L., \& Zheng, W. 2005, ApJ, 630,
  68

\begin{deluxetable}{cccccc}
\tabletypesize{\small} 
\tablecaption{LBG-2377 Components
\label{comp}} 
\tablewidth{0pt} 
\tablehead{\colhead{Comp.} & \colhead{$z_{Phot}$} & 
\colhead{$z_{ISM}$\tablenotemark{a}} & \colhead{$z_{Ly\alpha}$} &
\colhead{$\Delta v_{ISM}$\tablenotemark{a}\tablenotemark{b}} & 
\colhead{$\Delta v_{Ly\alpha}$\tablenotemark{b}}} 
\startdata 
A &  3.0385  &  3.0354  & (3.0488)\tablenotemark{c} &  -230  &  
(765)\tablenotemark{c}\\ 
B &  3.0343  &  3.0308  &  \ldots  &  -260  &   \ldots\\
C &  3.0289  &  3.0258  &  3.0416  &  -230  &   945 \\ 
D &  3.0244  &  3.0209  &  3.0341  &  -260  &   725 \\
E & (3.0161)\tablenotemark{c} & (3.0143)\tablenotemark{c} & 
(3.0263)\tablenotemark{c} & (-135)\tablenotemark{c} &  
(760)\tablenotemark{c}\\
\enddata
\tablenotetext{a}{Averaged fit to multiple features.}
\tablenotetext{b}{Approximate velocity offset with respect to the 
stellar photospheric velocities in km sec$^{-1}$.}
\tablenotetext{c}{Features are weak and respective values have a 
lower confidence level.}
\end{deluxetable}
\end{thebibliography}
\end{document}